\documentclass[11pt]{article}
  
\usepackage{amsmath,amssymb}

\usepackage{epsfig}  
\usepackage{graphicx}               
\usepackage{url}
\usepackage{hyperref}
\usepackage{slashed}
\usepackage[makeroom]{cancel}
\usepackage{cite}

\usepackage{pstricks}
\usepackage{color}

\usepackage{tikz}
\usetikzlibrary{positioning}
\usetikzlibrary{decorations.pathmorphing}
\usetikzlibrary{decorations.markings}
\usetikzlibrary{snakes}
 
\def\beq{\begin{equation}}
\def\eeq{\end{equation}}
\def\bea{\begin{eqnarray}}
\def\eea{\end{eqnarray}}
\def\nn{\nonumber}
\def\nl{\nonumber\\}

\def\roughly#1{\mathrel{\raise.3ex\hbox
{$#1$\kern-.75em\lower1ex\hbox{$\sim$}}}}

\def\lesssim{\mathrel{\hbox{\rlap{\hbox{\lower4pt\hbox{$\sim$}}}\hbox{$<$}}}}
\def\gtrsim{\mathrel{\hbox{\rlap{\hbox{\lower4pt\hbox{$\sim$}}}\hbox{$>$}}}}

\def\sla#1{\raise.15ex\hbox{$/$}\kern-.57em #1}

\def\barr{\begin{eqnarray}}
\def\earr{\end{eqnarray}}
\def\beast{\begin{eqnarray*}}
\def\eeast{\end{eqnarray*}}
\def\eea{\end{eqnarray}}
\def\bit{\begin{itemize}}

\def\nn{\nonumber}
\def\nl{\nonumber\\}
\def\roughly#1{\mathrel{\raise.3ex\hbox
{$#1$\kern-.75em\lower1ex\hbox{$\sim$}}}}

\def\lesssim{\mathrel{\hbox{\rlap
{\hbox{\lower4pt\hbox{$\sim$}}}\hbox{$<$}}}}
\def\gtrsim{\mathrel{\hbox{\rlap
{\hbox{\lower4pt\hbox{$\sim$}}}\hbox{$>$}}}}

\newcommand{\bcln}{b \to c l^- \bar{\nu}_{l}}

\newcommand{\barBstdtn}{\bar{B} \to D^{*} \tau \bar{\nu}_{\tau}}

\def\BDtaunu{\bar{B} \to D^+ \tau^{-} \bar{\nu_\tau}} 
 
\def\BDlnu{\bar{B} \to D^+ \ell^{-} \bar{\nu_\ell}}
\def\BDstartaunu{\bar{B} \to D^{*+} \tau^{-} \bar{\nu_\tau}}
\def\BDstarellnu{\bar{B} \to D^{*+} l^{-} \bar{\nu_\ell}}

\def\BDstarlnu{\bar{B} \to D^{*} \ell^{-} \bar{\nu_\ell}}

\def \thl {{\theta_l}}
\def \thD {{\theta_{D^*}}}

\def\A0{{\cal{A}}_{0}}
\def\Apar{{\cal{A}}_{\|}}
\def\Aperp{{\cal{A}}_{\perp}}

\def\AtP{{\cal{A}}_{tP}}

\def\ssV{ \sin {2 \theta_{D^*}}}
\def\sVsq{ \sin^2{\theta_{D^*}}}

\def\c2V{ \cos { 2 \theta_{D^*}}}
\def\cVsq{ \cos^2{\theta_{D^*}}}
\def\sl{ \sin { \theta_{l}}}
\def\s2l{ \sin {2 \theta_{l}}}
\def\slsq{ \sin^2{ \theta_{l}}}
\def\cl{ \cos { \theta_{l}}}
\def\c2l{ \cos { 2 \theta_{l}}}
\def\clsq{ \cos^2{ \theta_{l}}}

\def\ubar{\overline{u}}

\def\Qbar{\overline{Q}}

\def\nubar{{\overline{\nu}}}
\def\psibar{\overline{\psi}}

\def\L{\mathcal{L}}

\def\A{\mathcal{A}}

\title{  
\vspace*{-2.3cm}  
\begin{flushright}  
\normalsize{  
UMISS-HEP-2014-01
  }  
\end{flushright}  
\vspace{1.5cm}  
\Large  
\textbf{
The Azimuthal  $B \to D^{*} \tau^{-} \bar{\nu_\tau}$ Angular Distribution
with Tensor Operators
}\vspace*{1.0cm}   
}

\author{ Murugeswaran Duraisamy\,$^{a}$, Preet Sharma\, $^a$ and Alakabha Datta\,$^a$
\vspace{5mm}
\\
$^{a}$ \normalsize\emph{Department of Physics and Astronomy, University of Mississippi,Oxford, MS 38677-1848, USA}  
}

\date{}

\begin{document}  

\setcounter{page}{0}  
\maketitle  

\vspace*{1cm}  
\begin{abstract} 
In  a recent paper we performed a comprehensive study of the impact of   new-physics operators with different Lorentz structures
on $\BDstarellnu$ decays, $\ell = e^-, \mu^-, \tau^- $,
involving the $b \to c l \nu_\ell$ transition. In this work we extend the previous calculation by including tensor operators. In the case of $\BDstartaunu $, we present the full three angle and $q^2$ angular distribution
with tensor new physics operators with complex couplings. The impact of the tensor operators on various observables in the angular distribution, specially the azimuthal observables including the CP violating triple product asymmetries are discussed.
It is shown that these azimuthal observables are very useful in discriminating different new physics operators. 
Finally we consider  new physics leptoquark models with tensor interactions and show how the presence of additional scalar operators modify the predictions 
of the tensor operators.
\end{abstract}  
  
\thispagestyle{empty}  
\newpage  
  
\setcounter{page}{1}

\baselineskip18pt   

\vspace{-3cm}

\section{Introduction}
\label{sec:intro}
The search for new physics(NP) beyond the Standard Model (SM) of particle physics is going on at the energy frontier in colliders such as the LHC and at the intensity frontier at high luminosity experiments. In the intensity frontier, the B factories, BaBar and Belle, have produced an enormous  quantity of data and there is still a lot of data to be analyzed from both experiments. The LHCb and Belle II will continue
the search for NP through precision measurements in the b quark system.
There are a variety of ways in which NP in B decays can be observed \cite{dattaBnp}.
In this NP search, the second and third generation quarks and leptons may be  quite special because they are comparatively heavier and could be relatively more sensitive to NP. As an example, in certain versions of the two Higgs doublet models (2HDM) the couplings of the new Higgs bosons are proportional to the masses and so NP effects are more pronounced for the heavier generations. Moreover, the constraints on NP  involving, specially the third generation leptons and quarks, are somewhat weaker allowing for larger NP effects \cite{datta_thirdgen}. 

The semileptonic decays of B meson to the $\tau$ lepton is 
mediated by a $W$ boson in the SM and it is quite well understood 
theoretically. In many models of NP  this decay gets 
contributions from additional states like new vector bosons, leptoquarks or 
new scalar particles. These new states can affect the semileptonic $b \to c$ and
$b \to u$ transitions. 
The exclusive decays $\BDtaunu$ 
and $\BDstartaunu$ are important places to look for NP 
because, being three body decays, they offer a host of observables 
in the angular distributions of the final state particles. The 
theoretical uncertainties of the SM predictions have gone down 
significantly in recent years because of the developments in 
heavy-quark effective theory (HQET). The experimental situation 
has also improved a lot since the first observation of the decay 
$\BDstartaunu$ in 2007 by the Belle Collaboration 
\cite{Matyja:2007kt}. After 2007 many improved measurements have 
been reported by both the BaBar and Belle collaborations and the 
evidence for the decay  $\BDtaunu$ has also been found 
\cite{Aubert:2007dsa,Adachi:2009qg,Bozek:2010xy}. 
Recently, the BaBar collaboration with their full data sample 
of an integrated luminosity 426 fb$^{-1}$ has reported the measurements 
of the quantities \cite{Lees:2012xj,Lees:2013uzd}
\begin{eqnarray}
\label{babarnew}
R(D) &=& \frac{BR(\BDtaunu)}
{BR(\BDlnu)}=0.440 \pm 0.058 \pm 0.042\, ,
\nonumber \\
R(D^*) &=& \frac{BR(\BDstartaunu)}
{BR(\BDstarlnu)}=0.332 \pm 0.024 \pm 0.018 \, ,
\end{eqnarray}
where $l$ denotes the light lepton $(e, \mu)$. The SM predictions for $R(D)$ and $R(D^*)$ are 
\cite{Lees:2012xj,Fajfer:2012vx,Sakaki:2012ft}
\begin{eqnarray}
R(D) &=& 0.297 \pm 0.017 \, ,
\nonumber \\
R(D^*) &=& 0.252 \pm 0.003 \,,
\end{eqnarray}
which deviate from the BaBar measurements by 2$\sigma$ and 2.7$\sigma$ 
respectively. The BaBar collaboration themselves reported a 3.4$\sigma$ 
deviation from SM when the two measurements of Eq.~(\ref{babarnew}) are taken 
together.  In this work we do not include the Belle measurements in our average. 

These deviations could be sign of NP and already certain models of NP have been considered to explain the data \cite{Fajfer:2012vx,Fajfer:2012jt, Crivellin:2012ye,Datta:2012qk,Becirevic:2012jf, Deshpande:2012rr,  Celis:2012dk,Choudhury:2012hn,Tanaka:2012nw,Ko:2012sv,Fan:2013qz,Biancofiore:2013ki,Celis:2013jha,Duraisamy:2013pia,Dorsner:2013tla,Sakaki:2013bfa}.
In Ref.~\cite{Datta:2012qk}, we calculated various observables in $\BDtaunu$ and $\BDstartaunu$ decays with NP  using an effective Lagrangian approach.
The Lagrangian contains two quarks and two leptons  scalar, pseudoscalar, vector, axial vector and tensor operators. Considering  subsets of the NP operators at a time, the  coefficient of these operators can be fixed from the BaBar measurements and then one can  study the effect of these operators on the various observables.  In \cite{Duraisamy:2013pia}  we extended the work of Ref.~\cite{Datta:2012qk}
by providing the full angular distribution with NP. 
In particular we focused on the CP violating observables which are  the triple product (TP) asymmetries  \cite{TP}.
In the SM these TP's  vanish to a very good approximation as the decay is dominated by a single amplitude. Hence, non-zero measurements of these terms are clear signs of NP without any hadronic uncertainties. Note, in the presence of NP
with complex couplings the TP's are non-zero and  depend on the form factors.
Another probe of CP violation using the decay of the $\tau$ from $\BDtaunu$ to multipion decays was recently considered \cite{Hagiwara:2014tsa}.

In this work we include tensor operators in the NP effective Hamiltonian and study their effects on various observables, particularly focusing on the azimuthal observables, including the triple products. Tensor operators were discussed earlier for these decays in
\cite{Tanaka:2012nw,Biancofiore:2013ki,Dorsner:2013tla, Sakaki:2013bfa}.
In this work, for $\BDstartaunu $, we present the full three angle and $q^2$ angular distribution including tensor new physics operators with complex couplings. This represents the full angular distribution with the most general new physics.
In our calculations we focus on the effects of the tensor operators on 
observables that are sensitive to the azimuthal angle $\chi$ which is the angle between the decay plane of the $D^*$ meson and the off-shell $W^*$. The triple products are the term proportional to the $\sin\chi$ in the angular distribution.
For completeness we will also discuss other observables such as the $q^2$ differential
distribution as well as the polarization and forward-backward asymmetries.

Finally, we note that tensor operators are often accompanied by other operators in specific NP models. Hence as an example of tensor operators we  consider a leptoquark model that has both tensor and scalar operators. We  study how the presence of the scalar operators modify the predictions of the different observables in the angular distribution.

The paper is organized in the following manner. In Sec. 2   we set up our formalism where we introduce the effective Lagrangian for NP with tensor operators and define the various observables 
in 
$\BDstartaunu$ decays. In Sec. 3 we present an explicit leptoquark NP model
where we show how tensor operators may arise and consider a few cases.
In Sec. 4  we present the numerical predictions which include constraints on the NP couplings as well as predictions for the various observables with NP in $\BDstartaunu$. 
Finally in Sec. 5 
 summarize the results of our analysis.
\section{Formalism}
\label{formalism}

In the presence of NP, the effective Hamiltonian for the quark-level transition $\bcln$  can be written in the form \cite{ccLag} 
\bea
{\cal{H}}_{eff} &=& \frac{4 G_F V_{cb}}{\sqrt{2}} \Big[ (1 + V_L)\,[\bar{c} \gamma_\mu P_L b] ~ [\bar{l} \gamma^\mu P_L \nu_l] \, +  V_R \, [\bar{c} \gamma^\mu P_R b] ~ [\bar{l} \gamma_\mu P_L \nu_l] \nl && \, + S_L \, [\bar{c} P_L b] \,[\bar{l}  P_L \nu_l] \, +  S_R \, ~[\bar{c} P_R b] \,~ [\bar{l}  P_L \nu_l]  \,  + T_L \, [\bar{c} \sigma^{\mu \nu} P_L b] \,~[\bar{l} \sigma_{\mu \nu} P_L \nu_l]\Big]\,,
\eea 
where  $G_F = 1.1663787(6) \times 10^{-5} GeV^{-2}$ is the Fermi coupling constant, $V_{cb}$ is the Cabibbo-Kobayashi-Maskawa (CKM) matrix element, $P_{L,R} = ( 1 \mp \gamma_5)/2$  are the projectors of negative/positive chiralities. We use $\sigma_{\mu \nu} = i[\gamma_\mu, \gamma_\nu]/2$ and
assume the neutrino to be always left chiral.
Further, we do not assume any relation between $b \to u l^- \nu_l$ and $\bcln$ transitions and hence  do not include constraints from
$B \to \tau \nu_{\tau}$. 
The SM  effective Hamiltonian corresponds to $V_L = V_R = S_L = S_R =  T_L = 0$.

\subsection{$\BDstartaunu$  angular distribution}
\label{barBstdtn}
\begin{figure}[h!]
\centering
\includegraphics[width=9.5cm]{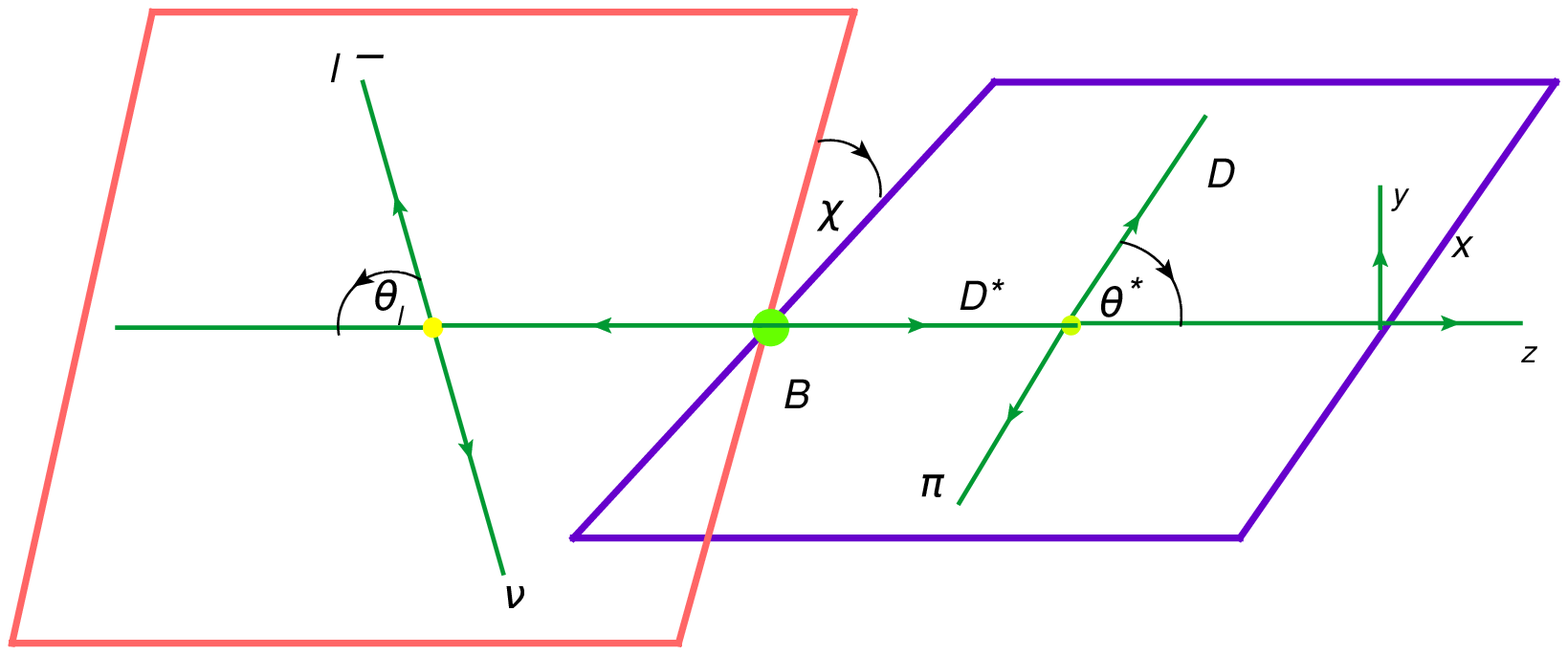}
      \caption{The description of the angles $\theta_{\l,D^{*}}$ and $\chi$ in the angular distribution of $\bar{B}\rightarrow D^{*}  (\rightarrow D \pi)l^-\nu_l$  decay.}
\label{fig-DstlnuAD}
\end{figure}

The complete three-angle distribution for the decay  $\bar{B}\rightarrow D^{*}  (\rightarrow D \pi)l^- \bar{\nu}_l$ in the presence of NP can be expressed in terms of four kinematic variables $q^2$, two polar angles $\theta_l$, $\theta_{D^*}$, and the azimuthal angle $\chi$. The angle $ \theta_l$ is the polar angle between the charged lepton and the direction opposite to the $D^*$ meson in the $(l \nu_l)$ rest frame. The angle $\theta_{D^*}$  is the polar angle between the D meson and the direction of the $D^*$ meson in the  $(D \pi)$ rest frame. The angle $ \chi$  is the azimuthal angle between the two decay planes spanned by the 3-momenta of the $(D \pi)$ and $(l \nu_l)$ systems. These angles are described in Fig.~\ref{fig-DstlnuAD}. The three-angle distribution can be obtained by using the helicity formalism:

\noindent We can write the angular distribution explicitly for easy comparison with previous literature \cite{Dungel:2010uk, Richman:1995wm,Korner:1989qb,Korner:1989qa}
in terms of the helicity amplitudes
\bea
\label{3-foldAD_ex}
\frac{d^4\Gamma}{dq^2\, d\cos\theta_l\, d\cos\theta_{D^*}\, d\chi} & = &
 \frac{9}{32 \pi} NF   \Big(\sum^8_{i = 1} I_i + \frac{m_l^2}{q^2} \sum^8_{j = 1} J_i \Big), \nl
\eea
where we can define the ${I_i}$ and ${J_i}$ as,
\bea
I_1 & = & 4 \cVsq \Big(\slsq |\A_0|^2 + 8 |A_{0T}|^2 \Big[1+\cos 2\thl \Big]\Big), \nl
J_1 & = & 4\cVsq\Big( \Big[ |\A_0|^2 \clsq + |\AtP|^2-2Re[\AtP\A_0^*]\cl \Big]\nl && +  4 \Big[|A_{0T}|^2 (1-\cos 2\thl)-(\frac{m_l^2}{q^2})^{-1/2} Re(\A_{0T}{\A_0}^*) \Big]\Big) ,\nl
I_2 & = & \sVsq \Big( \Big[ (|\Apar|^2 +|\Aperp|^2 )(1+\clsq)- 4Re[\Apar\Aperp^*]\cl \Big] \nl && +8\Big[  (|{\Apar}_T|^2+|{\Aperp}_T|^2 )(1-\clsq) \Big] \Big) ,\nl
J_2 &=& \sVsq\Big( \slsq (|\Apar|^2 +|\Aperp|^2 )+8\Big[(|{\Apar}_T|^2 +|{\Aperp}_T|^2 )(4+\clsq) \nl && -4 Re({\Apar}_T {{\Aperp}_T}^*)\sl  -2(\frac{m_l^2}{q^2})^{-1/2} {Re({\Apar}_T {\Apar}^* + {\Aperp}_T {\Aperp}^*)(1-\sl)} \Big] \Big), \nl
I_3 & = & -\sVsq \slsq \cos{2 \chi} \Big( [|\Apar|^2 -|\Aperp|^2]-16 [|{\Apar}_T|^2 -|{\Aperp}_T|^2 ]\Big),\nl
J_3 & = & \sVsq \slsq \cos{2 \chi} \Big([|\Apar|^2 -|\Aperp|^2 ] -16 (\frac{m_l^2}{ q^2})^{-1/2}[|{\Apar}_T|^2 -|{\Aperp}_T|^2 ]\Big),\nl 
I_4 & = & -2\sqrt{2} \ssV \sl \cos{\chi}  Re[\Aperp \A_0^*],  \nl
J_4 & = & 2\sqrt{2} \ssV \sl \cos{\chi}  \Big(Re[\Apar \AtP^*]-16 \Big[Re({\Aperp}_T A_{0T}^*) \nl && + (\frac{m_l^2}{q^2})^{-1/2}Re(A_{0T}{\Aperp}^* + {\Aperp}_T {\A_0}^* - {\Apar}_T A_{tP}^*) \Big]\Big),  \nl
I_5 & = & 2\sqrt{2} \ssV \sl \cl \cos{\chi} \Big(Re[\Apar \A_0^*]-16 Re[{\Apar}_T A_{0T}^*]\Big),  \nl
J_5 & = & -2\sqrt{2} \ssV \sl \cl \cos{\chi}  \Big(Re[\Apar \A_0^*]-16 [{\Apar}_T A_{0T}^*]\Big),  \nl
I_6 & = & 2\sin^2\thD \sin^2\thl \sin 2\chi Im[\Apar\Aperp^*],\nl
J_6 & = & -2\sin^2\thD \sin^2\thl \sin 2\chi Im[\Apar\Aperp^*],\nl
I_7 & = & -2 \sqrt{2}\sin 2\thD \sin\thl \sin\chi Im[\Apar\A_0^*],\nl
J_7 & = &  -2 \sqrt{2}\sin 2\thD \sin\thl \sin\chi\Big( Im[\Aperp\AtP^*] \nl && - 4 (\frac{m_l^2}{q^2})^{-1/2} Im(A_{0T}{\Apar}^* -{\Apar}_T A_{0}^* + {\Aperp}_T A_{tP}^* )\Big), \nl
I_8 & = & \sqrt{2}  \sin 2\thD \sin 2\thl \sin\chi Im[\Aperp\A_0^*], \nl
J_8 & = & -\sqrt{2}  \sin 2\thD \sin 2\thl \sin\chi Im[\Aperp\A_0^*]. \
\eea
The various helicity amplitudes are defined in the appendix.

It will be convenient to rewrite the angular distribution as \cite{Alok:2011gv}
\begin{align}
\label{eq:3-foldAD}
\frac{d^4\Gamma}{dq^2\, d\cos\theta_l\, d\cos\theta_{D^*}\, d\chi}=
 \frac{9}{32 \pi} NF  & \Bigg\lbrace \cos^2\thD \Big(V^0_{1} + V^0_{2}\cos 2\thl + 
     V^{0}_{3} \cos\thl \Big) \nl & + 
 \sin^2\thD \Big(V^{T}_{1} + V^{T}_{2} \cos 2\thl + 
 V^{T}_{3}\cos\thl \Big) \nl & + 
   V^{T}_{4} \sin^2\thD \sin^2\thl  \cos 2\chi + 
   V_1^{0T} \sin 2\thD \sin 2\thl   \cos\chi  \nl &+ 
   V_2^{0T}  \sin 2\thD \sin\thl  \cos\chi + 
   V^{T}_{5} \sin^2\thD \sin^2\thl \sin 2\chi \nl &+ 
   V_3^{0T}  \sin 2\thD \sin\thl \sin\chi + 
   V_4^{0T}  \sin 2\thD \sin 2\thl \sin\chi\Bigg\rbrace \; ,
   \end{align}
where the quantity $N_F$ is
\bea
\label{NF}
N_F &=& \Big[ \frac{G^2_F |p_{D^*}| |V_{cb}|^2 q^2 }{3\times 2^{6}\pi^3 m^2_B} \Big(1-\frac{m_l^2}{q^2}\Big)^2 ~Br(D^*\rightarrow D\pi)\Big] \; .
\eea
The momentum of the $D^{*}$ meson in the B meson rest frame is denoted as $|p_{D^{*}}|=\lambda^{1/2}_{}(m^2_B,m^2_{D^{(*)}},q^2)/2 m_B$  with  $\lambda(a,b,c) = a^2 + b^2 + c^2 - 2 (ab + bc + ca)$.  

The twelve angular coefficients ($V_i$) depend on the couplings, kinematic variables and form factors, and are given in terms of $\barBstdtn$ helicity amplitudes in  appendix.
 We use HQET to expand the form factors in terms of certain parameters, which are then fixed from the angular distribution for $B \to D^{*} \ell^{-} \bar{\nu_\ell}$, where $\ell= e , \mu $ \cite{Dungel:2010uk}. Our basis assumption is that
$B \to D^{*} \ell^{-} \bar{\nu_\ell}$ decays are described by the SM.

The following single-differential angular distributions allow access to various  observables that can be used to probe for NP. 
The differential decay rate  $d\Gamma/dq^2$ can be obtained after performing integration over all the angles 
\bea
\label{eq1:DBRDstsq}
\frac{d\Gamma}{dq^2 } &=& \frac{3 N_F}{4} 
(A_{L}+A_{T}) \,.
\eea
Here the $D^*$ meson's longitudinal and transverse polarization amplitudes  $A_L$ and $A_T$ are
\bea
\label{HL}
A_L &=& \Big(V_1^{0}  - \frac{1}{3} V_2^{0}  \Big),\quad A_T = 2 \Big(V_1^{T}  - \frac{1}{3} V_2^{T} \Big) ~.
\eea
Furthermore, one can also explore the $q^2$ dependent of ratio
\bea
\label{eq:RDst}
R_{D^*}(q^2) &=&\frac{d Br[\BDstartaunu]/dq^2 }{d Br[\BDstarlnu]/dq^2}\,.
\eea

 By integrating out the polar angles $\theta_l$, $\theta_{D^*}$, and the azimuthal angle $\chi$ in different kinematic regions, various 2-fold angular distributions can be obtained. For a detailed discussions see our previous work \cite{Duraisamy:2013pia}. Here, we have updated these angular distributions with the new tensor couplings. Our results agree with the corresponding angular distributions in \cite{Sakaki:2013bfa}. Several observables can be defined through the 2-fold angular distributions. The $D^*$ polarization fraction $F_L$, the forward-backward asymmetry $A_{FB}$  for the leptons, the  azimuthal asymmetries, including the three transverse asymmetries $A_C^{(1, 2, 3)}$, and the three T-odd  CP asymmetries  $A_T^{(1, 2, 3)}$,  are defined in terms of angular coefficients $V_i's$ \cite{Duraisamy:2013pia}:
\begin{align}
\label{eq:obs1}
F^{D^*}_L(q^2) &= \frac{A_L}{A_L +A_T }\, \hspace*{2cm} A^{D^*}_{FB}(q^2) = \frac{ V^{T}_3 + \frac{1}{2} V^{0}_3}{A_L + A_T},\ \nn\\   
A^{(1)}_C (q^2)& = \frac{4 V^T_4}{3 (A_L+ A_T)} \, \hspace*{2cm}A^{(1)}_T  (q^2)= \frac{4 V^T_5}{3(A_L+ A_T)},\ \nn\\   
A^{(2)}_C (q^2)&= \frac{ V^{0T}_2}{ (A_L+ A_T)} \, \hspace*{2cm}
A^{(2)}_T(q^2) = \frac{ V^{0T}_3}{(A_L+ A_T)},\, \nn\\
A^{(3)}_C (q^2)&= \frac{ V^{0T}_1}{ (A_L+ A_T)}  \, \hspace*{2cm}  
A^{(3)}_T(q^2) =\frac{ V^{0T}_4}{(A_L+ A_T)}.
\end{align}.


In closing this section we note that even though we are focused on the $\BDstartaunu$ decay  the $\BDtaunu$ decay is used to constrain the
NP operators. The $\BDtaunu$  angular distribution, with tensor operators, can be written as, 
\bea
\label{eq:DDRBdtn}
\hspace*{-2.5cm}\frac{d \Gamma^D}{dq^2 d\cos{\theta_l}}&=& 2 N_D |p_D|~ \Bigg[  |H_0|^2  \sin^2{\theta_l}  +  \frac{m^2_l}{q^2} (H_0 \cos{\theta_l}-H_{tS})^2 \nl && \hspace*{-2.5cm} + 8 \Bigg(\Big((1 + \frac{m_l^2}{q^2}) + (1 -  \frac{m_l^2}{q^2}) \cos{2 \theta_l}\Big) |H_T |^2   -\frac{ m_l }{\sqrt{q^2}} Re[H_T (H^*_0 - H^*_{tS}\cos{\theta_l})]\Bigg) \Bigg]\,,
\eea
where  the prefactor  $N_D = \frac{G^2_F |V_{cb}|^2 q^2}{256 \pi^3 m^2_B} \Big(1-\frac{m_l^2}{q^2}\Big)^2$. The helicity amplitudes are
\bea
\label{eqApp3:BDAmp}
H_0 &=& \sqrt{\frac{\lambda_D}{q^2}} (1 + g_V )  F_+(q^2)\,,\quad
H_t = \frac{m^2_B -m^2_D}{\sqrt{q^2}}   (1 + g_V )   F_0(q^2)\,,\nl
H_S &=&  -\frac{m^2_B -m^2_D}{m_b (\mu) - m_c(\mu)}\, g_S \, F_0(q^2)  \,,\quad
H_T = -\frac{\sqrt{\lambda_D}}{m_B + m_D} \,T_L \,F_T(q^2)\,,
\eea
where $g_{V,A} =  V_R \pm V_L$ and $g_{S,P} =  S_R \pm S_L$. 
In addition, the $H_t$ and the $H_S$ amplitudes arise in the combination,
\bea
H_{tS} &=& \Big(H_t - \frac{\sqrt{q^2}}{m_\tau} H_S \Big)\,.
\eea
{{\bf The results in eq.{(\ref{eq:DDRBdtn})}  agree with the $\BDtaunu$  angular distribution in \cite{Sakaki:2013bfa}. }}

\section{ An Explicit Model}
Many extensions of the SM, motivated by a unified description of quarks and leptons, predict the existence of new scalar and vector bosons, called leptoquarks, which decay into a quark and a lepton. These particles carry nonzero baryon and lepton numbers, color and fractional electric charges. 
The most general dimension four  $SU(3)_c\times SU(2)_L\times U(1)_Y$ invariant Lagrangian of  leptoquarks satisfying baryon and lepton number conservation was considered in Ref~\cite{Buchmuller:1986zs}.
As the tensor operators in the effective Lagrangian get contributions only from scalar leptoquarks, we will focus only on scalar leptoquarks and consider the case where the leptoquark is a weak doublet or a weak singlet. The weak doublet leptoquark, $R_2$
has the quantum numbers   $(3,2,7/6)$ under  $SU(3)_c\times SU(2)_L\times U(1)_Y$
while the singlet leptoquark $S_1$ has the quantum numbers $(\bar{3},1,1/3)$.
 
The interaction Lagrangian that induces contributions to the $b \to c \ell \overline{\nu}$ process is \cite{Tanaka:2012nw}
\bea
\L_{2}^{\rm LQ} & =&  \left( g_{2L}^{ij}\,\ubar_{iR}R_2^T L_{jL} + g_{2R}^{ij}\,\Qbar_{iL} i\sigma_2 \ell_{jR} R_2 \right), \nonumber\\
\L_{0}^{\rm LQ} & =& \left( g_{1L}^{ij},\Qbar_{iL}^c i\sigma_2 L_{jL} + g_{1R}^{ij},\ubar_{iR}^c \ell_{jR} \right)S_1, \
\label{leptoquark_lag}
\eea
where $Q_i$ and $L_j$ are the left-handed quark and lepton $SU(2)_L$ doublets respectively, while $u_{iR}$, $d_{iR}$ and $\ell_{jR}$ are the right-handed up, down quark and charged lepton $SU(2)_L$ singlets. Indices $i$ and $j$ denote the generations of quarks and leptons, and $\psi^c = C\psibar^T=C\gamma^0\psi^*$ is a charge-conjugated fermion field.
The fermion fields are given in the gauge eigenstate basis
and one should make the transformation to the mass basis. Assuming the quark mixing matrices to be hierarchical, and considering only the leading contribution we can ignore the effect of mixing.
After performing the Fierz transformations, one finds the general Wilson coefficients  at the leptoquark mass scale  contributing to the $b \to c \tau \nubar_l$ process:
\bea
   \label{LQ_WC}
      S_L &=& { 1 \over 2\sqrt2 G_F V_{cb} }  \left[ -{g_{1L}^{33}g_{1R}^{23*} \over 2M_{S_1}^2} - {g_{2L}^{23}g_{2R}^{33*} \over 2M_{R_2}^2} \right], \nonumber\\
      T_L &=& { 1 \over 2\sqrt2 G_F V_{cb} }  \left[ {g_{1L}^{33}g_{1R}^{23*} \over 8M_{S_1}^2} - {g_{2L}^{23}g_{2R}^{33*} \over 8M_{R_2}^2} \right]. \,
\eea
It is clear from Eq.~(\ref{LQ_WC}) that the weak singlet leptoquark and the weak  doublet can add constructively
or destructively to the Wilson's coefficients of the scalar and tensor operators in the effective Hamiltonian. We can now consider various scenarios. In the first case the singlet and the doublet
scalar leptoquark couplings are such that the scalar operator couplings are enhanced and the tensor operator couplings are suppressed. This scenarios has been studied before \cite{Datta:2012qk,Duraisamy:2013pia}. 
Hence, the first case, called Case. (a), we will study is when
 the tensor operators is enhanced and the scalar operator suppressed.
The results of the pure tensor coupling are presented in the next section.

In this section we will also consider the possibilities where both the scalar and the tensor operators are present and are of similar sizes. In the most general case both
the singlet and doublet leptoquarks are present and so both
 the scalar and tensor operators  appear in the effective Hamiltonian.
As there is limited experimental information, including both the singlet and the doublet leptoquarks will allow us more flexibility in fitting for the Wilson's coefficients
but this will come with the price of less precise predictions for the various observables.
We can, therefore, consider the simpler cases when only a singlet or a doublet leptoquark are present. In these cases, from Eq.~(\ref{LQ_WC}) the coefficients of scalar operators and the tensor operators have the same magnitudes. One can now consider two further cases: 

Case. (b): In this case only the weak doublet scalar leptoquark  $R_2$
 is present.
It was shown recently \cite{Arnold:2013cva} that this is one of the two
 minimal renormalizable scalar leptoquark model, where the standard model is augmented only by one additional scalar representation of $SU(3) \times SU(2) \times U(1)$ and which  do  not allow proton decay at the tree level.

The relations between the scalar and tensor couplings in Eq.\ref{LQ_WC} are valid at the leptoquark mass scale, $m_{\rm LQ}$. We have to run them down to the $b$ quark mass scale using
the scale dependence of the scalar and tensor currents at leading logarithm approximation
\begin{equation}
      S_L(\mu_b) = \left[ \alpha_s(m_t) \over \alpha_s(\mu_b) \right]^{\gamma_S \over 2\beta_0^{(5)}} \left[ \alpha_s(m_{\rm LQ}) \over \alpha_s(m_t) \right]^{\gamma_S \over 2\beta_0^{(6)}} S_L(m_{\rm LQ}) \,, \\ 
      T_L(\mu_b) = \left[ \alpha_s(m_t) \over \alpha_s(\mu_b) \right]^{\gamma_T \over 2\beta_0^{(5)}} \left[ \alpha_s(m_{\rm LQ}) \over \alpha_s(m_t) \right]^{\gamma_T \over 2\beta_0^{(6)}} T_L(m_{\rm LQ}) \,,
   \label{eq:QCDrunning}
\end{equation}
where the anomalous dimensions of the scalar and tensor operators are $\gamma_S=-6C_F=-8$, $\gamma_T=2C_F=8/3$ respectively and $\beta_0^{(f)}=11-2n_f/3$ \cite{Dorsner:2013tla}. 
Choosing a value for the leptoquark mass we can run the couplings to the  $b$-quark scale which is chosen to be $\mu_b=\overline m_b=4.2$~GeV.

In the simplified scenario with the presence of only one type of leptoquark, namely $R_2$ or $S_1$, the scalar $S_L$ and tensor $T_L$ Wilson coefficients are no longer independent: one finds that at the scale of leptoquark mass, $ m_{\rm LQ}$,  $S_L(m_{\rm LQ})=\pm T_L(m_{\rm LQ})$. Then, using Eq.~\eqref{eq:QCDrunning}, one obtains the relation at the bottom mass scale,
\begin{equation}
   S_L (\overline m_b) \simeq \pm7.8 \, T_L (\overline m_b) \,.
\end{equation}
for a leptoquark mass of 1 TeV \cite{Tanaka:2012nw}.

It is interesting to note that the same coupling that appears in the process $b \to c \tau \nubar_l$
also appears in the $ t \to c \tau^+ \tau^-$ decay and if the
components of the doublet leptoquark have the same mass,
  then we can have a prediction for this decay based on data from $B \to D^{(*)} \tau \overline{\nu}_\tau$ transition.

Case. (c): In this case only the singlet leptoquark is present and the relevant Wilson's coefficients can be obtained from Eq.\ref{LQ_WC}.

\section{Numerical analysis }
The model independent and dependent  numerical results for the various observables in the angular distribution of $\BDstartaunu$  decay are discussed in this section.
\subsection{Model independent results}
For the numerical calculation, we use the $B \to D$ and $B \to D^*$ form factors in the heavy quark effective theory(HQET) framework \cite{Neubert:1993mb,Caprini:1997mu}. A detailed discussions on the $B \to D^*$  and $B \to D$ form factors and their numerical values can be found in \cite{Sakaki:2013bfa}. The constraints on the complex NP couplings in  the  $\bcln$ effective Hamiltonian come  from  the measured $R(D)$ and $R(D^*)$ in Eq.~(\ref{babarnew}) at 95\% C.L. We  vary the free parameters in the HQET form factors  within their error bars. All the other numerical values are taken from \cite{Nakamura:2010zzi} and \cite{Asner:2010qj}. The allowed  ranges for the  NP couplings are then used for predicting the possible allowed ranges  for the observables.

It is important to point out that the combination of  couplings $g_V = V_R + V_L$ appears  in both $R(D)$ and $R(D^*)$, while $g_A = V_R - V_L$   appears only in $R(D^*)$. $V_R$ and $V_L$ receive constraints from both  $R(D)$ and $ R(D^*)$. While, the combination of couplings $g_S = S_R + S_L$ appears  only  in $R(D)$, $ g_P = S_R - S_L$  appears only in $R(D^*)$.   If NP is established in both $R(D)$ and $ R(D^*)$ then the cases of pure $g_A$ or $g_S$ or $g_P$ coupling are ruled out.  A detailed discussions on the effects of vector and scalar couplings on the various observables in the decays $\BDstarlnu$ and $\BDlnu$ can be found in our previous works \cite{Datta:2012qk,Duraisamy:2013pia}.

\begin{figure}[h!]
\centering
\includegraphics[width=6.5cm]{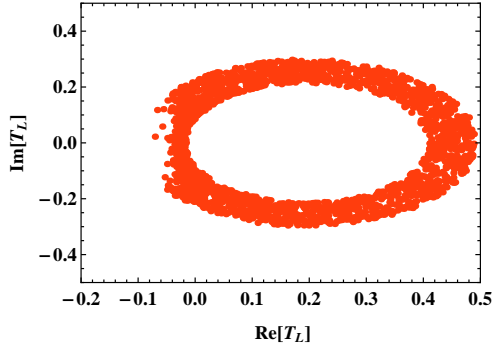}
      \caption{The allowed region for the complex coupling $T_L$ for Case. (a)  at 95\% C.L. }
\label{fig-TLConst}
\end{figure}

We first consider the Case. (a)  of the previous section where only the NP tensor operator is present in the effective Hamiltonian. In Fig.~(\ref{fig-TLConst}), the constraint on the parameter space of the pure tensor coupling  by both $R(D)$ and $R(D^*)$ measurements at 95\% C.L. is shown.  We find that the magnitude of tensor coupling satisfies  $|T_L| < 0.5$. 

\begin{figure}[h!]
\centering
\includegraphics[width=7.5cm]{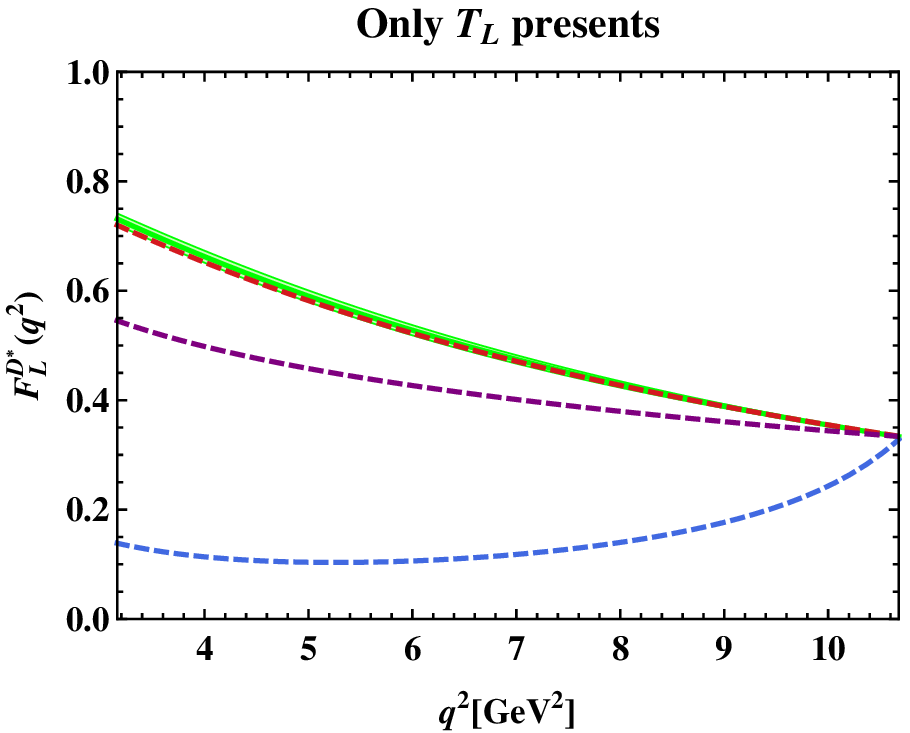} \includegraphics[width=7.5cm]{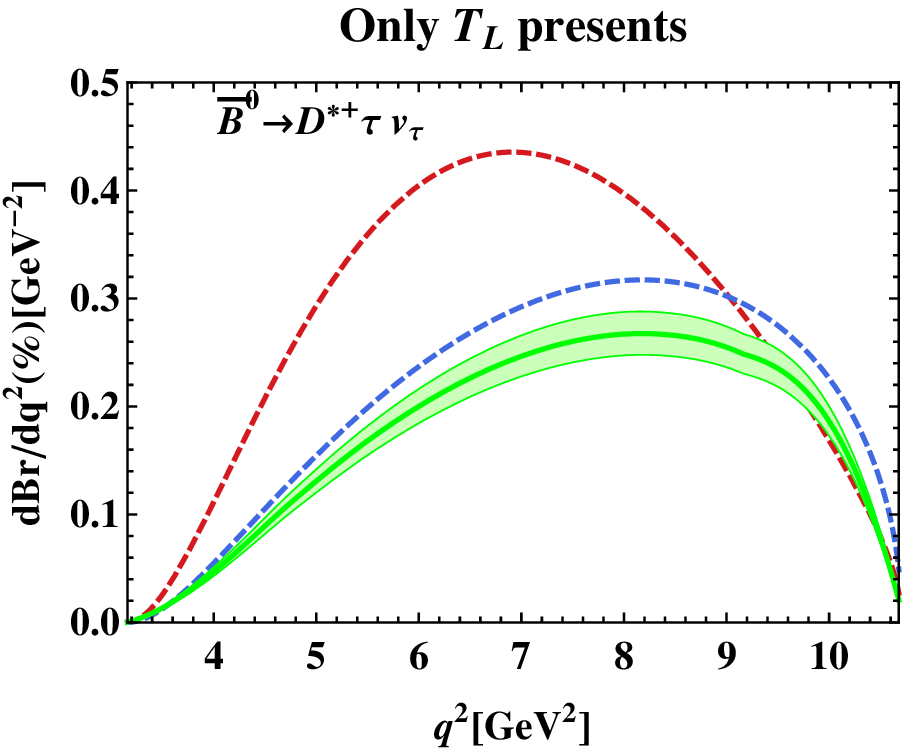}\\
\includegraphics[width=7.5cm]{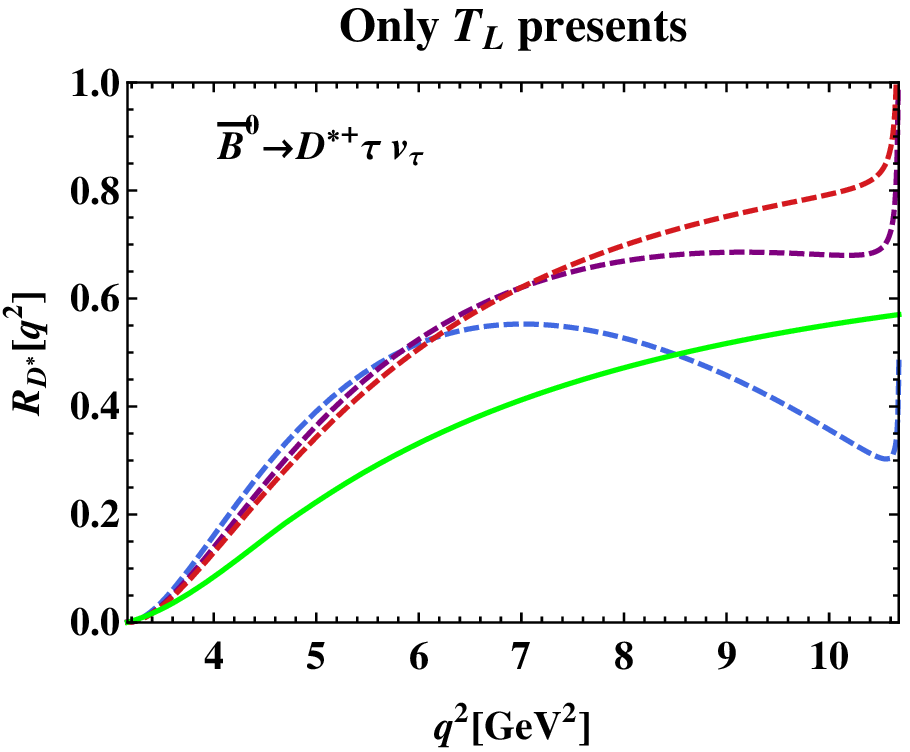} \includegraphics[width=7.5cm]{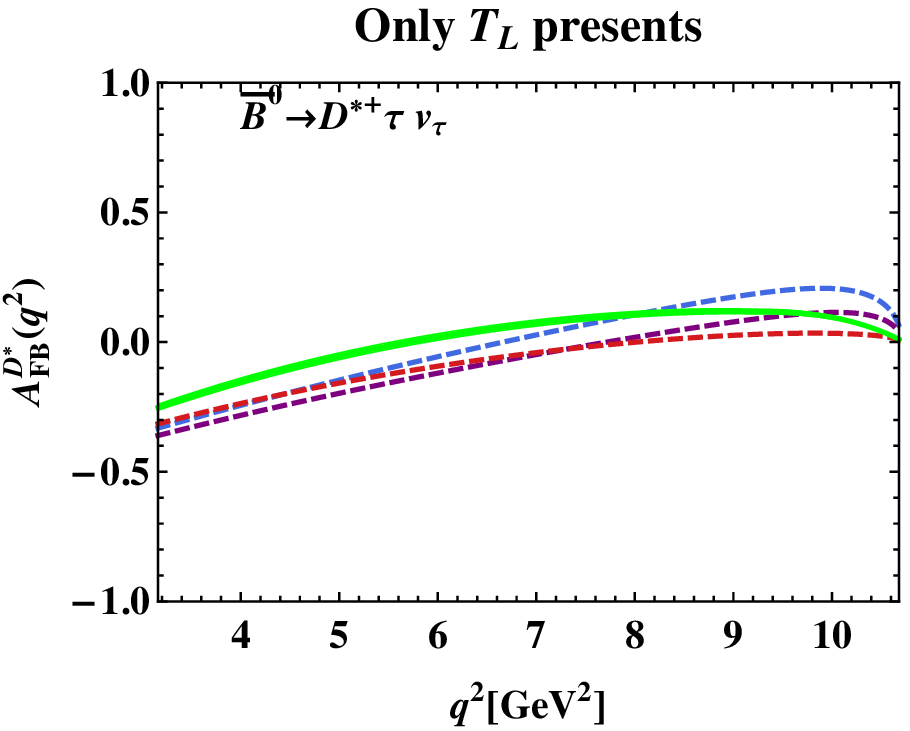}
      \caption{The predictions for the observables $F_{L}^{D^*}(q^2)$, differential branching ratio, $R_{D^*}(q^2)$, and $A_{FB}^{D^*}(q^2)$  for the decay $\bar{B}^0 \to D^{*+} \tau \nu_\tau$ in the presence of only $T_L$ coupling. The green  band corresponds to the SM prediction and its uncertainties. The values of the coupling  $T_L$ are chosen to show the maximum and minimum deviations from the SM expectations. }
\label{fig-FLdBRRDstAFBwithTL}
\end{figure}

The predictions for the differential branching ratio (DBR), $F^{D^*}_L(q^2)$,  $R(D^*)(q^2)$ and $A^{D^*}_{FB}(q^2)$ are shown in  Fig.~\ref{fig-FLdBRRDstAFBwithTL} for the allowed values of tensor coupling. It is clear that, the  DBR,  $F^{D^*}_L(q^2)$, and  $R(D^*)(q^2)$  get  considerable deviation from their SM expectation in this new physics scenario. The contribution of pure tensor coupling to the forward-backward asymmetry is of the order of $m_{\tau}/\sqrt{q^2}$, and  $A^{D^*}_{FB}(q^2)$ behaves similar to its SM expectation. 

\label{barBstdtn_Num}

\begin{figure}[h!]
\centering
\includegraphics[width=7.5cm]{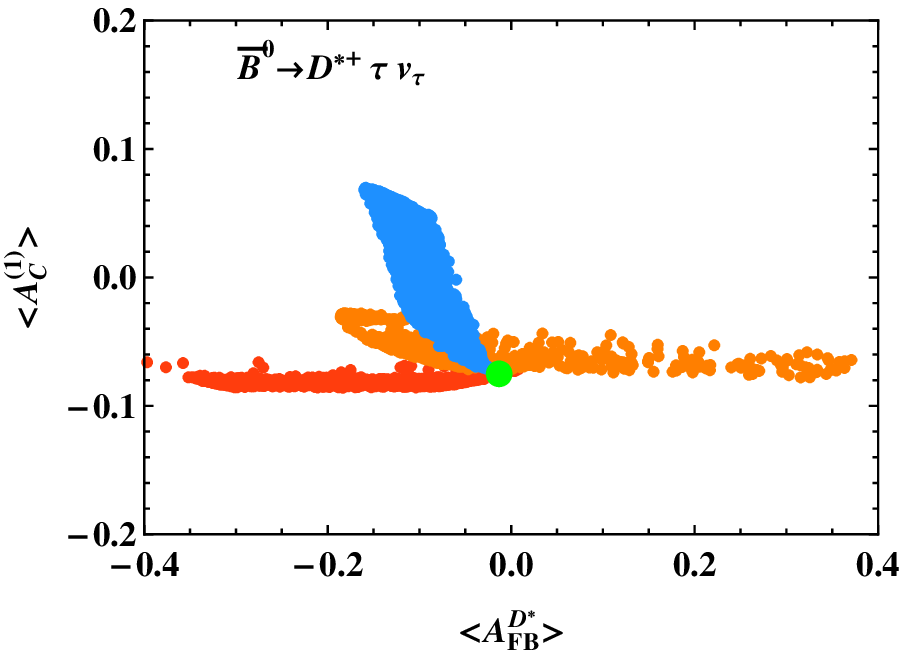}\includegraphics[width=7.5cm]{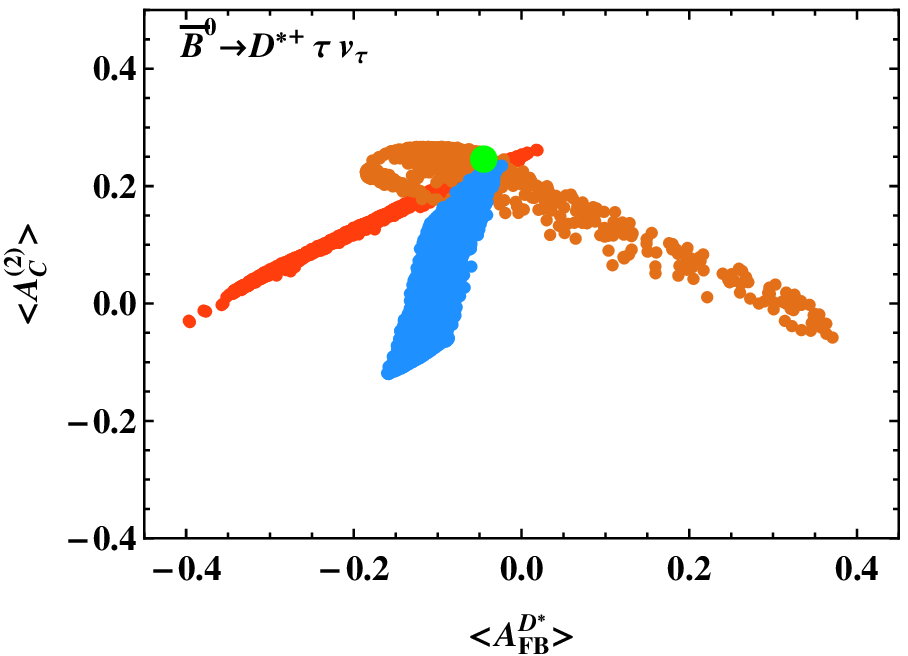}\\
\includegraphics[width=7.5cm]{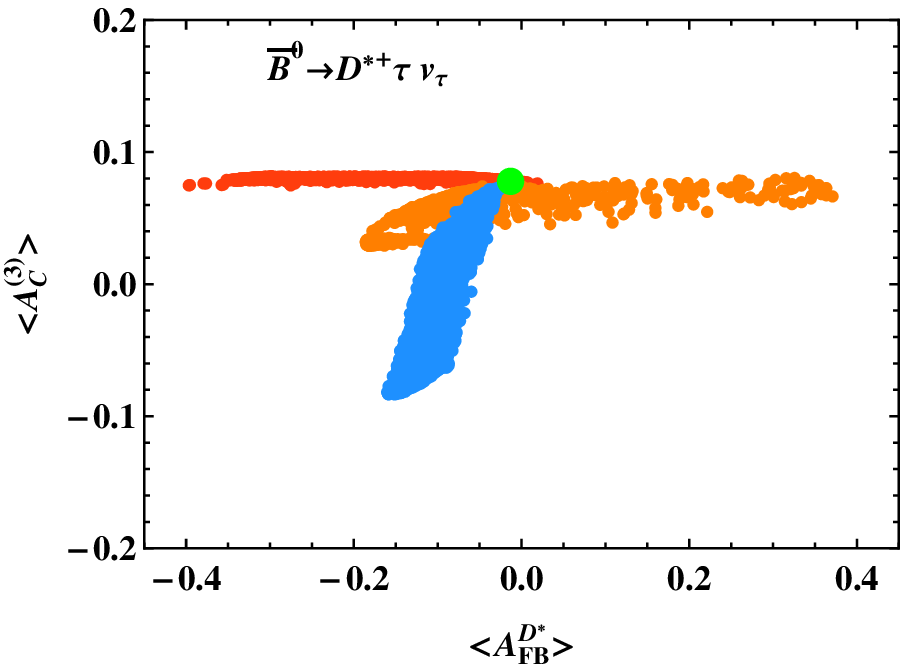}\includegraphics[width=7.5cm]{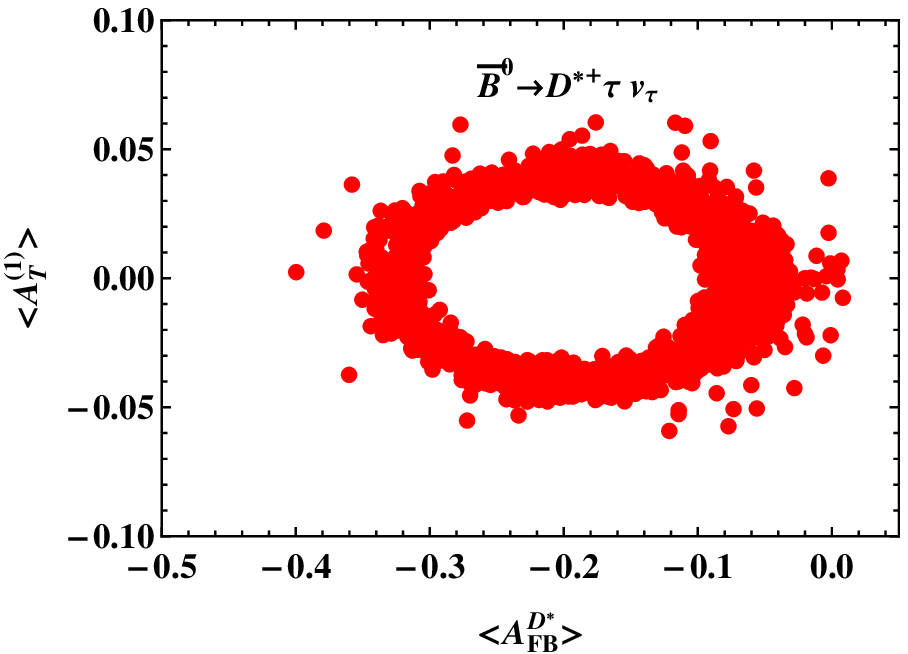}\\
\includegraphics[width=7.5cm]{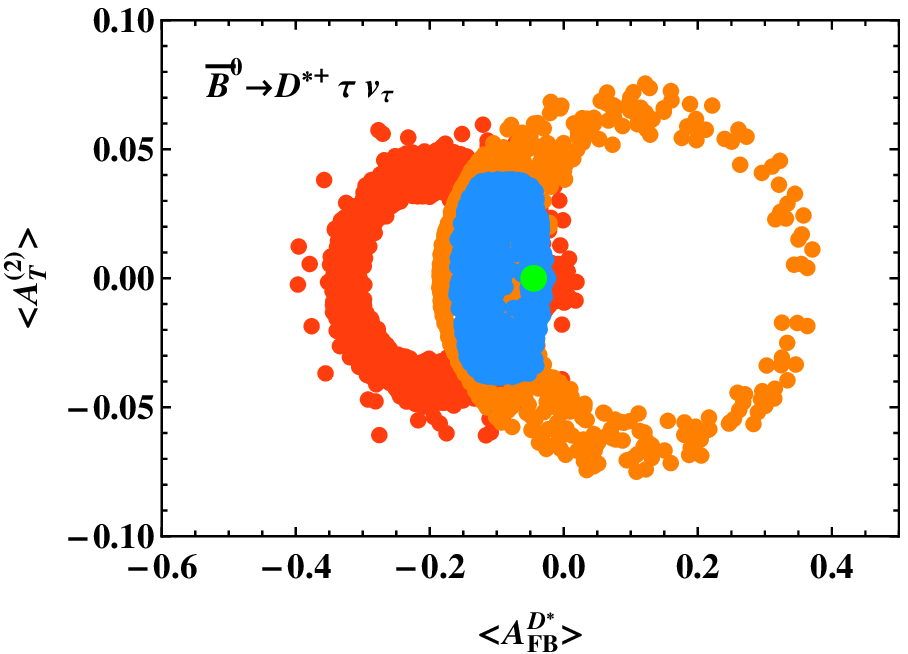}\includegraphics[width=7.5cm]{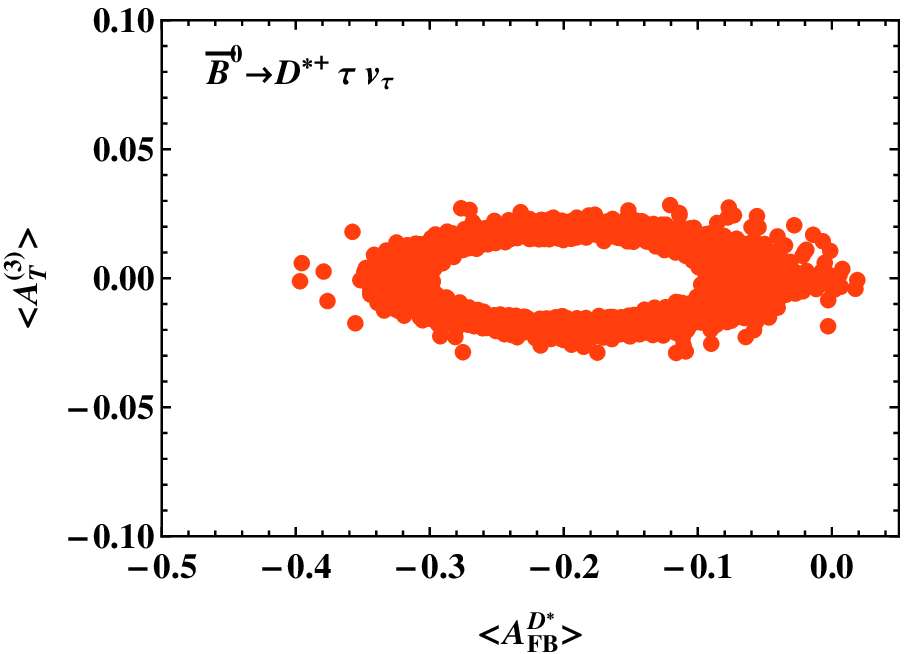}\\
      \caption{ The correlation plots between  $<A_C^{(1, 2, 3)}>$ (  $<A_T^{(1,2,3)}>$ )  and $<A^{D^*}_{FB}>$ in the presence of complex NP couplings. The red, orange and blue  scatter points   correspond to pure vector NP couplings $(V_L, V_R)$, pure scalar NP couplings $(S_L, S_R)$ , and  pure tensor NP coupling $(T_L)$. The scatter points are allowed by measurements of $R_D$ and $R_{D^*}$  at 95\% C.L. The green points correspond to the SM predictions for these quantities. }
\label{fig-ObsMInd}
\end{figure}

We now wish to analyze the sensitivity of the $q^2$-integrated azimuthal symmetries on the new tensor coupling,  and we present correlations of these  symmetries with respect to  the integrated forward-backward asymmetry (FBA).
The $q^2$-integrated FBA $<A^{D^*}_{FB}>$, the  three transverse asymmetries $ <A_C^{(1, 2, 3)}>$, and the three T-odd  CP asymmetries  $<A_T^{(1, 2, 3)}>$ can be obtained by separately integrating out the $q^2$-dependence in  the numerator and denominator of these quantities as expressed in Eq.(\ref{eq:obs1}). The panels of  Fig.(\ref{fig-ObsMInd}) show the correlation between the above six $q^2$-integrated asymmetries and $<A_{FB}>$ for the decay $\bar{B}^0 \to D^{*+} \tau \nu_\tau$. Note that, in this plot we also include predictions for the vector and scalar NP couplings. In each cases, the NP couplings satisfy the current measurements of $R_D$ and $R_{D^*}$  at 95\% C.L.   It is clear from these plots that  $<A^{D^*}_{FB}>$, and  $ <A_C^{(1, 2, 3)}>$ get considerable deviations from their SM expectation once we include the NP couplings. The T-odd CP asymmetry  $<A_T^{(2)}>$ is sensitive to all NP couplings, and is strongly correlated with   $<A^{D^*}_{FB}>$. The scalar NP couplings can enhance  this asymmetry about 5\% from its SM value. On the other hand,  $<A_T^{(1)}>$ and $<A_T^{(3)}>$ are only sensitive  to the vector couplings. These asymmetries are also strongly correlated with $<A^{D^*}_{FB}>$ in the presence of  vector NP couplings, and can  be enhanced up to 3\% from its SM value. Hence, the predictions  for  $<A^{D^*}_{FB}>$ and azimuthal symmetries have varying sensitivities to the different NP scenarios and these observables can be powerful probes of the structure of NP.

\subsection{Leptoquark model results}
\begin{figure}[h!]
\centering
\includegraphics[width=7.5cm]{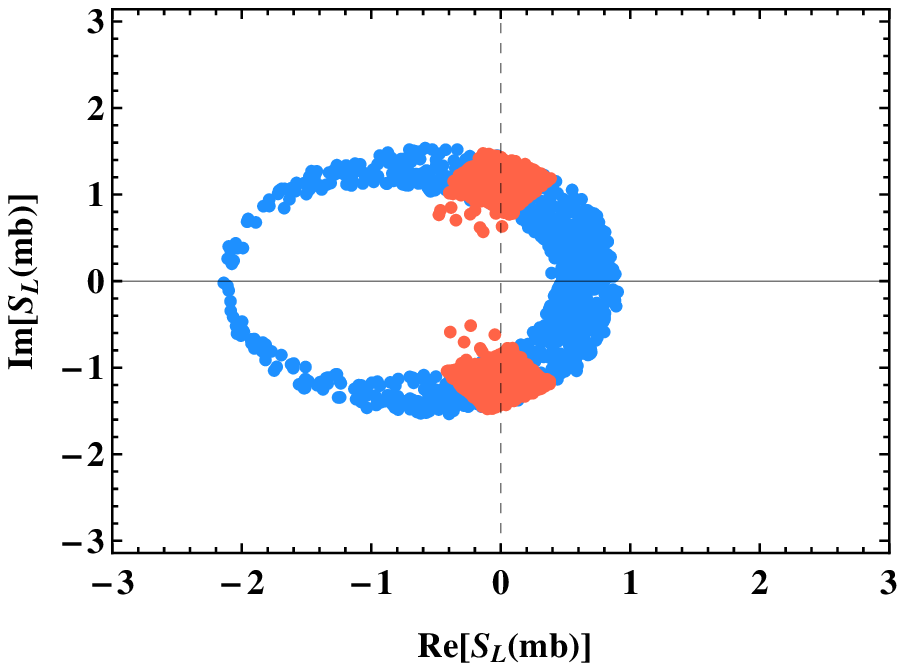}\includegraphics[width=7.5cm]{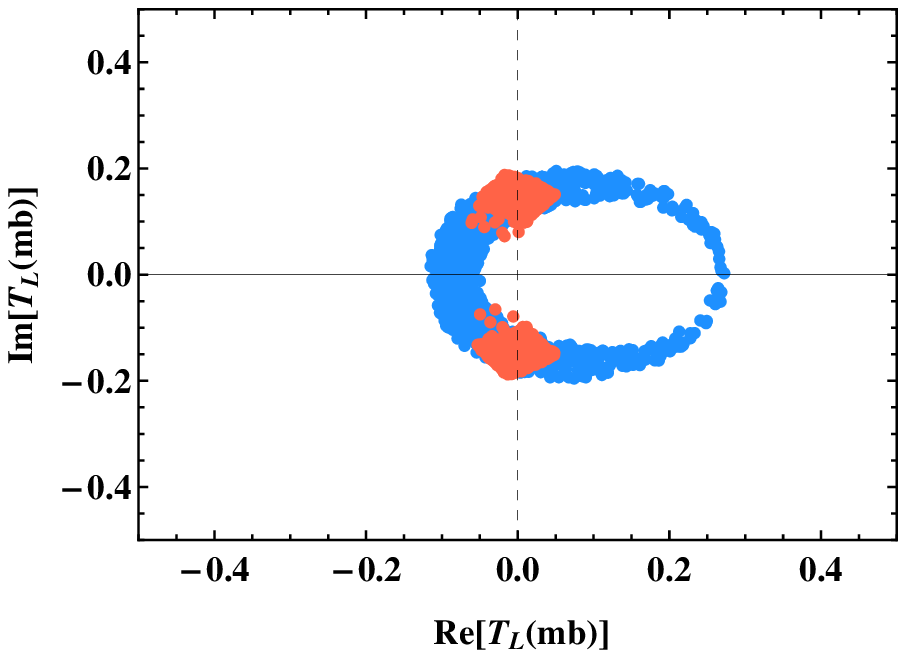}
      \caption{ The allowed regions for the leptoquark effective couplings $S_L$ and $T_L$ at $\mu_b = 4.2 GeV$.  The constraints on these NP couplings are from the measured $R(D)$ and $R(D^*)$ within the  $2 \sigma$ level. The red (blue)  scatter points  correspond to $ S_1 (R_2 )$  
 leptoquark models.}
\label{fig-SLTLconst}
\end{figure}

We next move to   Case.(b) and Case.(c) for the leptoquark with the mass scale of the order of 1 TeV. The allowed ranges for the leptoquark couplings at $\mu = m_b$
from the measured $R(D)$ and $R(D^*)$ values within the  $2 \sigma$ level
  are shown in  Fig.~(\ref{fig-SLTLconst}).
These results suggest that the magnitudes of the doublet and singlet leptoquark effective couplings, $g_{2L}^{23}g_{2R}^{33*} $ and  $g_{1L}^{33}g_{1R}^{23*}$ are  of ${\it{O}}(1)$. A similar conclusion is  obtained in \cite{Sakaki:2013bfa}.

\begin{figure}[h!]
\centering
\includegraphics[width=7.5cm]{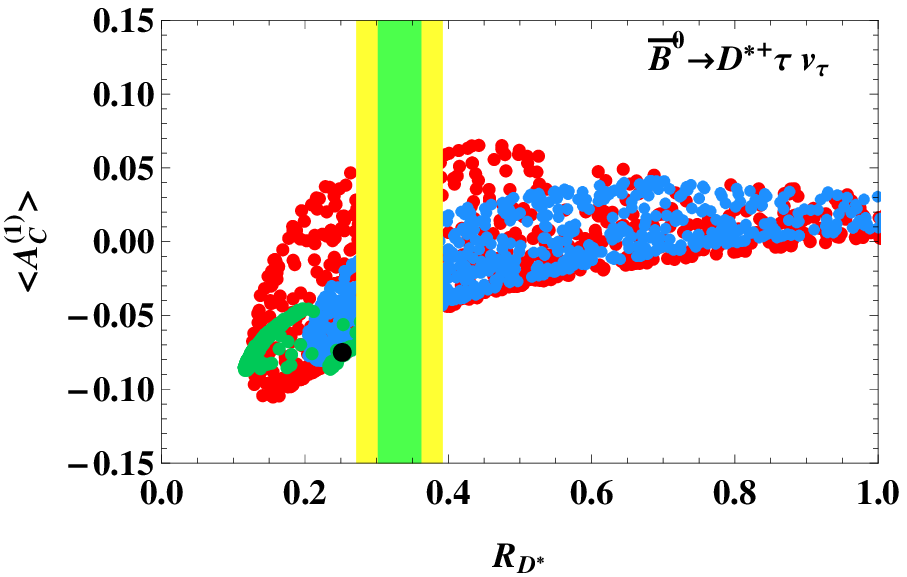}\includegraphics[width=7.5cm]{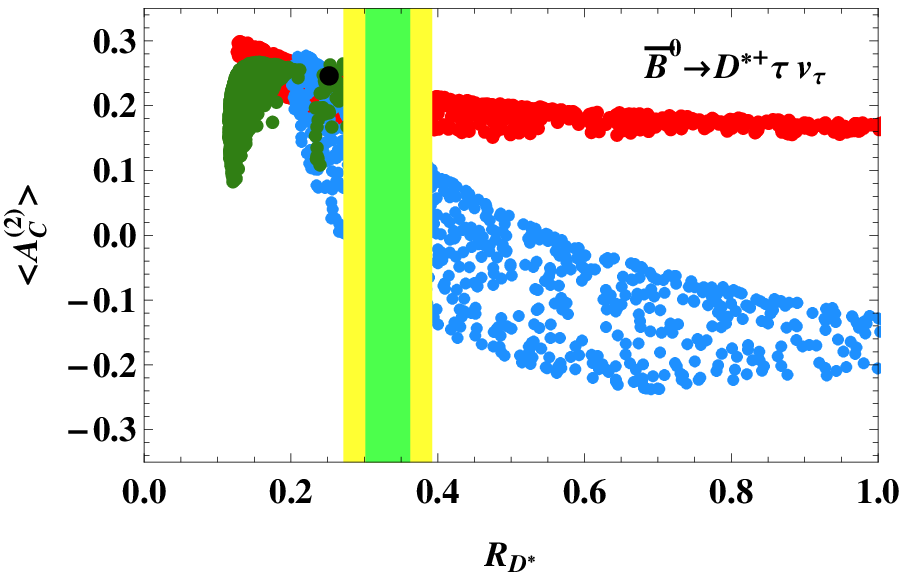}\\
\includegraphics[width=7.5cm]{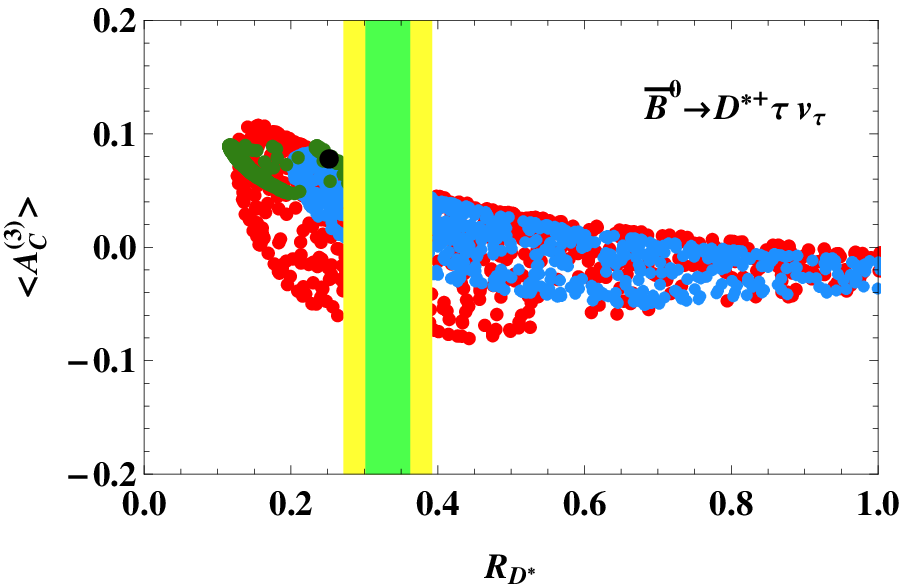}\includegraphics[width=7.5cm]{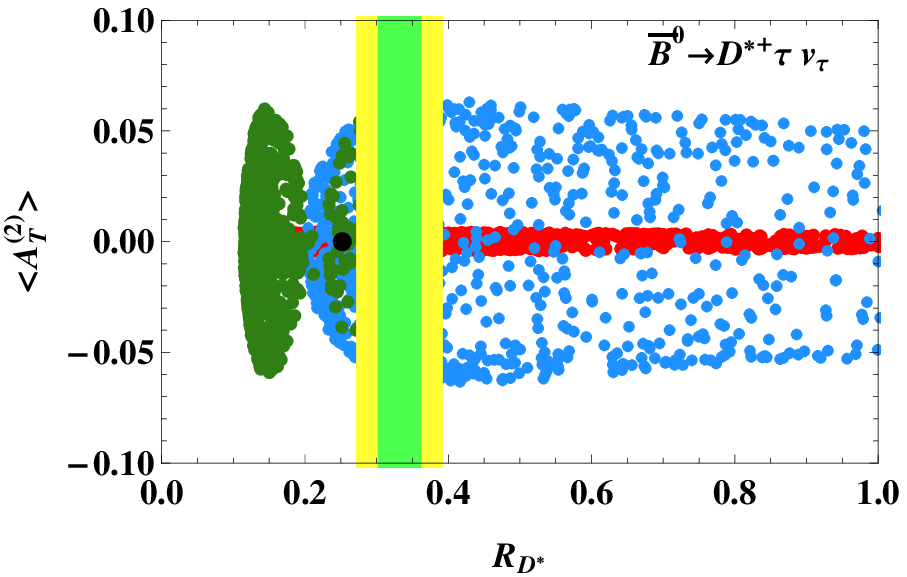}\\
      \caption{ The correlations between  $<A_C^{(1, 2, 3)}>$ (  $<A_T^{(2)}>$ )  and $R_{D^*}$ for three different NP scenarios: only $S_L$ coupling (green), $R_2$ leptoquark  coupling (red), and $S_1$ leptoquark  coupling (blue). The black points correspond to the SM predictions for these quantities. The vertical bands correspond to  $R_{D^*}$ data with  $\pm 1 \sigma$ (green) or $\pm 2 \sigma$ (yellow) errors. }
\label{fig-AsyvsRDst}
\end{figure}

The correlations between the  asymmetries $<A_C^{(1, 2, 3)}>$ and  $<A_T^{(2)}>$ and $R_{D^*}$  are shown in Fig.~({\ref{fig-AsyvsRDst}})  for three different NP scenarios: only $S_L$, only  $R_2$ leptoquark  ($S_L  =  7.8 T_L$),  and only $S_1$ leptoquark ($S_L  = -7.8 T_L$). These results imply that  $<A_C^{(1, 2, 3)}>$ and  $<A_T^{(2)}>$ can get sizeable contributions from the leptoquarks within the measured region of $R_{D^*}$.  It is interesting to note that the behavior of  $<A_C^{(2)}>$ is   different for  $R_2$ and $S_1$ leptoquark couplings. Hence this observable can be used to discriminate between the singlet and the doublet leptoquark models.

\begin{figure}[h!]
\centering
\includegraphics[width=7.5cm]{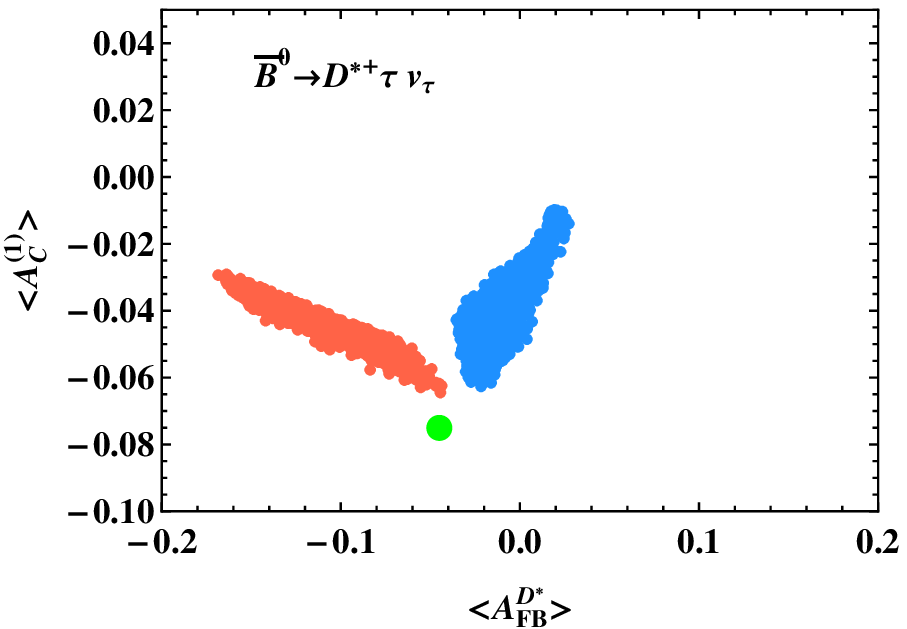}\includegraphics[width=7.5cm]{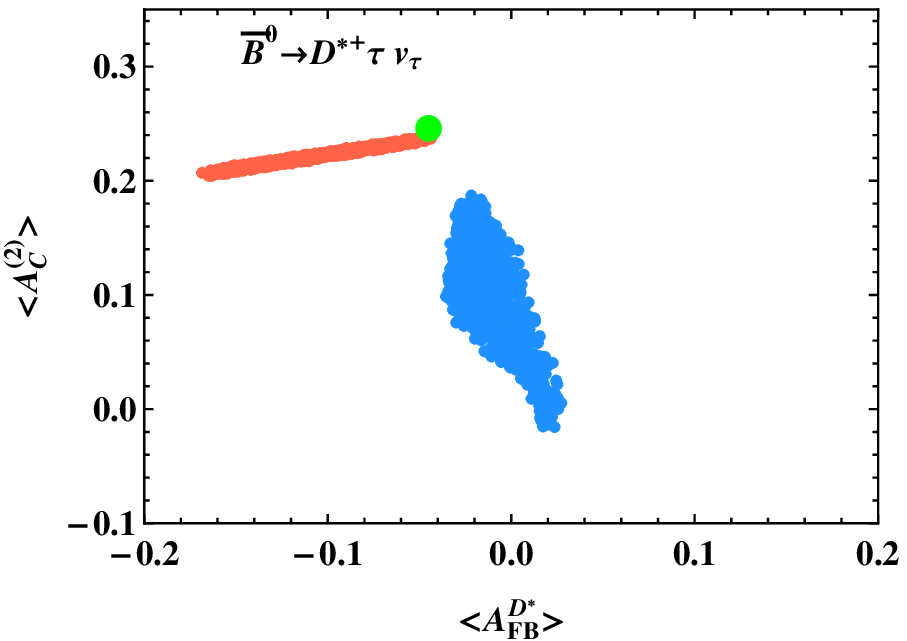}\\
\includegraphics[width=7.5cm]{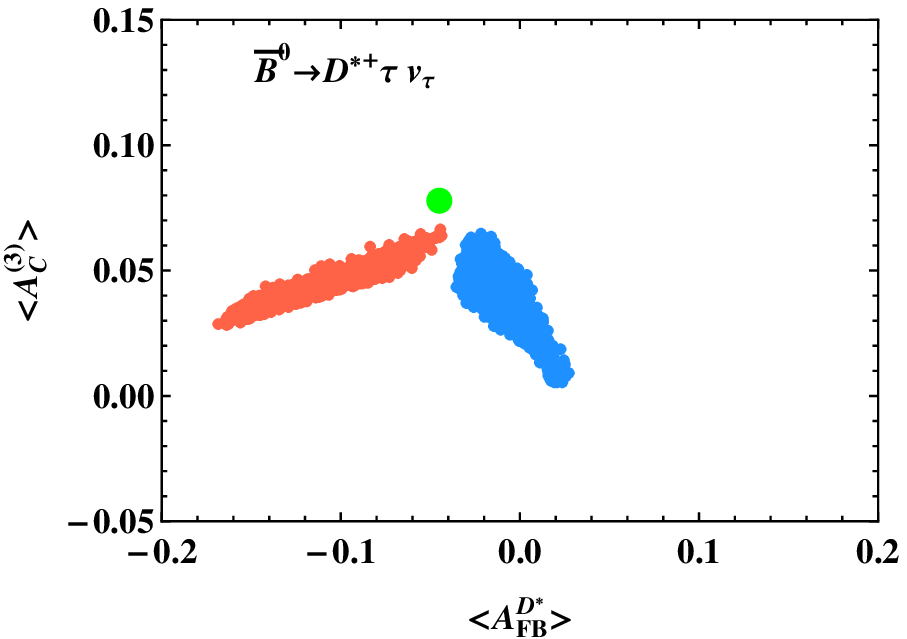}\includegraphics[width=7.5cm]{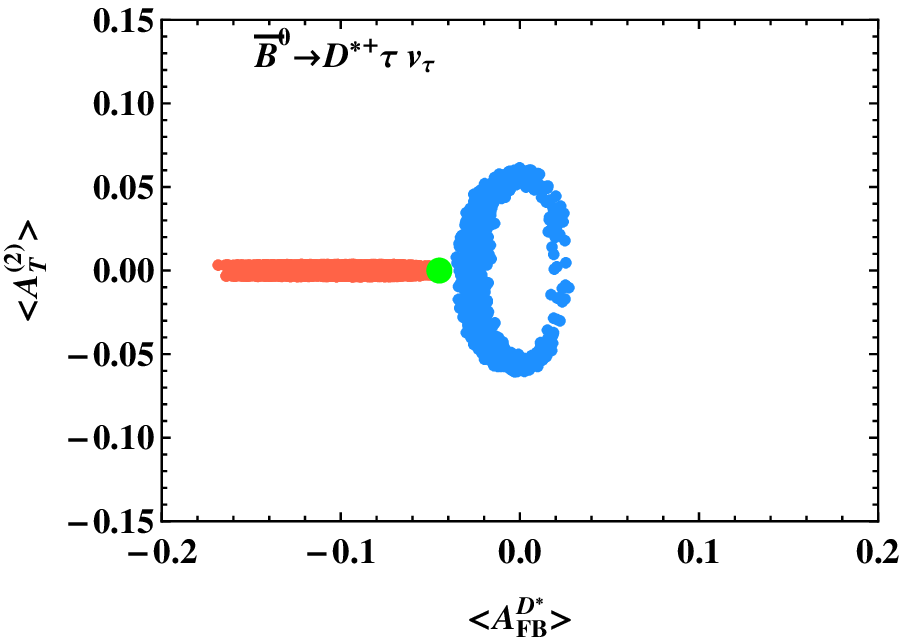}\\
      \caption{ The correlation plots between  $<A_C^{(1, 2, 3)}>$,  $<A_T^{(2)}>$,  and $<A^{D^*}_{FB}>$ in the presence of leptoquark contributions. The red (blue)  scatter points   correspond to $ R_2 (S_1 )$ leptoquarks. These scatter points satisfy the current measurements of $R_{D}$ and $R_{D^*}$ within the 2 $\sigma$ level. The green points in each panel correspond to the SM predictions for these quantities. }
\label{fig-ObsLQ}
\end{figure}

In Fig.(\ref{fig-ObsLQ}) we plot the correlations of $<A_C^{(1, 2, 3)}>$  and   $<A_T^{(2)}>$  with $<A^{D^*}_{FB}>$ in the presence of  $R_2$ and $S_1$ leptoquark contributions.  In each case, the constraints on the leptoquark couplings at $\mu= m_b$ are from the current measurements of $R_{D}$ and $R_{D^*}$ within the 2 $\sigma$ level. As in the case of pure tensor couplings, these plots show that the different leptoquark models produce very different predictions for the azimuthal asymmetries and so these observables can be very sensitive in ruling out different leptoquark models.

\section{Discussion and Summary}
\label{summary}
In summary we have discussed the effects of tensor operators in the decay $\BDstartaunu$
motivated by recent measurements which show deviation from the SM predictions in $\BDstartaunu$ and $\BDtaunu$. In this work we have presented the angular distribution for $\BDstartaunu$ with the most general new physics structure including tensor operators. We have then discussed the effects of the tensor operators on various observables that can be constructed out of the angular distribution. Our focus was on the azimuthal observables which include the important CP violating triple product asymmetries. We found that these azimuthal asymmetries, integrated over $q^2$, have different sensitivities to different NP structures and hence they can be powerful probes of the nature of the NP. These asymmetries also show strong correlations with the $q^2$ integrated forward-backward asymmetry. Tensor operators naturally arise in scalar leptoquark models and are accompanied by other scalar operators. We considered two leptoquark models
where the leptoquarks are weak singlets and doublets. We discussed  the predictions for the azimuthal observables in these models and found that these observables are very efficient in discriminating between the two leptoquark models. In particular we found that there is cancellation between the scalar and tensor components in the scalar doublet leptoquark model for one of the triple product asymmetries while this is not the case for the scalar singlet leptoquark model.

\section*{Acknowledgements} This work was supported in part by the National Science
Foundation under Grant No.\ NSF PHY-1068052.

\begin{appendix}
\section{Angular coefficients}
The twelve angular coefficients $V^{\lambda}_i$ in the $B\rightarrow D^{*}  (\rightarrow D \pi)l^- \bar{\nu}_l$ angular distribution depend on the couplings, kinematic variables and form factors. The expressions for these coefficients  are given in terms of the   hadronic helicity amplitudes of  the $\barBstdtn$ decay and summarized  according to the $D^*$ helicity combinations $\lambda_1 \lambda_2$:

The longitudinal $V^0$'s ($\lambda_1 \lambda_2 = 0 0$) are given by

\bea
\label{eq:V0}
V^0_1 &=&  2 \Big[\Big(1 +  \frac{m^2_l}{q^2} \Big) (| {\cal{A}}_0 |^2 + 16 | {\cal{A}}_{0T} |^2) + 
     \frac{2 m^2_l}{q^2} |{\cal{A}}_{tP}|^2 - \frac{ 16 m_l }{\sqrt{q^2}} ~{\rm Re}[{\cal{A}}_{0T} {\cal{A}}^*_0]\Big]~,\nl
V^0_2 &=& 2 \Big(1 -  \frac{m^2_l}{q^2} \Big) \Big[-| {\cal{A}}_0 |^2 + 16 | {\cal{A}}_{0T} |^2 \Big] ~,\nl
V^0_3 &=&  -8  {\rm Re}[ \frac{m^2_l}{q^2} {\cal{A}}_{tP}  {\cal{A}}^*_0 - \frac{ 4 m_l }{\sqrt{q^2}} {\cal{A}}_{tP} {\cal{A}}^*_{0T}]  ~.
\eea
The transverse $V^T$'s ($\lambda_1 \lambda_2 =++,--,+-,-+$) are given by

\bea
 \label{eq:VT}
V^T_1 &=& \Big[\frac{1}{2} \Big(3 +  \frac{m^2_l}{q^2} \Big) \Big(|{\cal{A}}_{\|}|^2 + |{\cal{A}}_\perp|^2 \Big) + 8 \Big(1 + \frac{3 m_l^2}{q^2} \Big) (|{\cal{A}}_{ \| T}|^2 + |{\cal{A}}_{\perp T}|^2)  -\frac{ 16 m_l }{\sqrt{q^2}}~{\rm Re}[{\cal{A}}_{ \| T} A^*_{\|} + {\cal{A}}_{\perp T} A^*_\perp]~\Big]~,\nl
V^T_2 &=&   \Big(1 -  \frac{m^2_l}{q^2} \Big)\Big[\frac{1}{2} \Big(|{\cal{A}}_{\|}|^2 + |{\cal{A}}_\perp|^2 \Big)- 8(|{\cal{A}}_{ \| T}|^2 + |{\cal{A}}_{\perp T}|^2) \Big]~,\nl
V^T_3 &=& 4  {\rm Re}\Big[ - {\cal{A}}_{\|}  A^*_\perp - \frac{ 16 m^2_l }{q^2} {\cal{A}}_{ \| T} {\cal{A}}^*_{\perp T} + \frac{ 4 m_l }{\sqrt{q^2}}~({\cal{A}}_{\perp T} {\cal{A}}^*_{\|} + {\cal{A}}_{ \| T} {\cal{A}}^*_\perp)  \Big] ~,\nl
V^T_4 &=&  \Big(1 -  \frac{m^2_l}{q^2} \Big) \Big[- \Big(|{\cal{A}}_{\|}|^2 - |{\cal{A}}_\perp|^2 \Big) + 
    16 (|{\cal{A}}_{ \| T}|^2 - |{\cal{A}}_{\perp T}|^2)\Big]~,\nl
V^T_5 &=&  2 \Big(1 -  \frac{m^2_l}{q^2} \Big) {\rm Im}[{\cal{A}}_{\|}  {\cal{A}}^*_\perp] ~.    
\eea

The mixed $V^{0T}$'s ($\lambda_1 \lambda_2 = 0\pm,\pm0$) are given by
\bea
\label{eq:VLT}
V^{0T}_1 &=& \sqrt{2}  \Big(1 -  \frac{m^2_l}{q^2} \Big) {\rm Re}[{\cal{A}}_{\|}  {\cal{A}}^*_0 -16 {\cal{A}}_{ \| T} {\cal{A}}^*_{0T} ] ~,\nl
V^{0T}_2 &=& 2 \sqrt{2} {\rm Re}\Big[- {\cal{A}}_{\perp}  {\cal{A}}^*_0 + \frac{m^2_l}{q^2} \Big( {\cal{A}}_{\|}  {\cal{A}}^*_{tP} - 16 {\cal{A}}_{\perp T}  {\cal{A}}^*_{0T}  \Big)  + \frac{ 4 m_l }{\sqrt{q^2}}~ \Big( {\cal{A}}_{0T} {\cal{A}}^*_{\perp} + {\cal{A}}_{\perp T} {\cal{A}}^*_0 - {\cal{A}}_{ \| T} {\cal{A}}^*_{tP} \Big)\Big]~,\nl
V^{0T}_3 &=&  2 \sqrt{2} {\rm Im}\Big[- {\cal{A}}_{\|}  {\cal{A}}^*_0 + \frac{m^2_l}{q^2}  {\cal{A}}_{\perp}  {\cal{A}}^*_{tP}   + \frac{ 4 m_l }{\sqrt{q^2}}~({\cal{A}}_{0T}{\cal{A}}^*_{\|} - {\cal{A}}_{ \| T} {\cal{A}}^*_0  +   {\cal{A}}_{\perp T}  {\cal{A}}^*_{tP}   ) \Big]~,\nl
V^{0T}_4 &=& \sqrt{2}  \Big(1 -  \frac{m^2_l}{q^2} \Big) {\rm Im}[{\cal{A}}_{\perp}  {\cal{A}}^*_0]~.
\eea

The expressions for the  hadronic helicity amplitudes  can be found in terms of  form factors for the $B \to D^*$ matrix elements \cite{Beneke:2000wa}

\bea
\label{tran_amp}
{\cal{A}}_0  &=&\frac{ (m_B + m_{D^*} )}{2 m_{D^*} \sqrt{q^2}} \Big[(m_B^2 - m_{D^*}^2 - q^2) A_1(q^2) -\frac{\lambda_{D^*}}{(m_B +m_{D^*})^2}  A_2(q^2) \Big](1 - g_A) \,,\nl
{\cal{A}}_{\pm}  &=& \Big[(m_B + m_{D^*})A_1(q^2) (1 - g_A) \mp \frac{\sqrt{\lambda_{D^*}}}{(m_B + m_{D^*})}V(q^2)(1 + g_V)\Big] \,,\nl
{\cal{A}}_{t}  &=& \frac{\sqrt{\lambda_{D^*}} }{\sqrt{q^2}} A_0(q^2) (1 - g_A) \,,\nl
{\cal{A}}_{P}  &=& \frac{\sqrt{\lambda_{D^*}}  }{ (m_b(\mu) + m_c(\mu))} A_0(q^2) g_P \,,\nl
{\cal{A}}_{0T}  &=& \frac{T_L}{2 m_{D^*} } \Big[(m_B^2 + 3 m_{D^*}^2 - q^2)T_2(q^2) -\frac{\lambda_{D^*}}{m_B^2 - m_{D^*}^2} T_3(q^2) \Big]\,,\nl
{\cal{A}}_{\pm T}  &=& T_L\Big[\frac{m_B^2 - m_{D^*}^2}{\sqrt{ q^2} }T_2(q^2) \pm \sqrt{\frac{\lambda_{D^*}}{q^2}} T_1(q^2) \Big]\,.\nl
\eea

 The $t$ and the $P$ amplitudes arise in the combination
\bea
\label{tp_comb}
{\cal{A}}_{tP} &=& \Big({\cal{A}}_t + \frac{\sqrt{q^2}}{m_\tau} {\cal{A}}_P \Big)\,.
\eea

Further, we define the transversity amplitudes ${\cal{A}}_{\|(T)}$ and ${\cal{A}}_{\perp (T)} $  in terms of the helicity amplitudes ${\cal{A}}_{\pm (T)}$  as

\bea
\label{tran_basis}
{\cal{A}}_{\| (T)} &= &\frac{1}{\sqrt{2}}  \left( {\cal{A}}_{+ (+T)} + {\cal{A}}_ {-(-T)}\right), \nonumber\\
{\cal{A}}_{\perp (T)} &= &\frac{1}{\sqrt{2}}  \left({\cal{A}}_{+ (+T)} - {\cal{A}}_{-(-T)} \right).
\eea
The expressions for the form factors $A_1(q^2)$, $A_2(q^2)$, $ A_0 (q^2)$, $V(q^2) $, $T_1(q^2)$, $T_2(q^2)$, and $T_3(q^2)$  in the heavy quark effective theory can be found in \cite{Falk:1992wt,Sakaki:2013bfa}.

\end{appendix}

\providecommand{\href}[2]{#2}\begingroup\raggedright\endgroup

\end{document}